\documentclass[10pt,conference]{IEEEtran}
\usepackage[english]{babel}
\usepackage{graphicx}
\usepackage{amsmath}
\usepackage{amsfonts}
\usepackage{amssymb}
\usepackage{amsfonts,eucal,bm}
\usepackage{a4wide}

\newtheorem{lemma}{Lemma}

\newtheorem{corollary}{Corollary}
\newtheorem{example}{Example}

\begin{document}

\title{A Smart Approach for  GPT Cryptosystem Based on Rank Codes}

\author{\authorblockN{Haitham Rashwan }
\authorblockA{Department of Communications \\
InfoLab21, South Drive\\
Lancaster University\\
Lancaster UK LA1 4WA\\
Email: h.rashwan@lancaster.ac.uk}
\and
\authorblockN{Ernst M. Gabidulin}
\authorblockA{Department of Radio Engineering\\
Moscow Institute \\
of Physics and Technology \\
(State University)\\
141700 Dolgoprudny,  Russia\\
Email: gab@mail.mipt.ru }
\and
\authorblockN{Bahram Honary}
\authorblockA{Department of Communications \\
InfoLab21, South Drive\\
Lancaster University\\
Lancaster UK LA1 4WA\\
Email: b.honary@lancaster.ac.uk }
}
\maketitle

\begin{abstract}
\noindent The concept of Public- key cryptosystem was innovated by McEliece's cryptosystem. The public key cryptosystem based on rank codes was presented in 1991 by Gabidulin --Paramonov--Trejtakov (GPT). The use of rank codes in cryptographic applications is advantageous since it is practically impossible to utilize combinatoric decoding. This has enabled using public keys of a smaller size. Respective structural attacks against this system were proposed by Gibson and recently by Overbeck. Overbeck's attacks break many versions of the GPT cryptosystem and are turned out to be either polynomial or exponential depending on parameters of the cryptosystem. In this paper, we introduce a new approach,  called the Smart approach, which is based on a proper choice of the distortion matrix $\mathbf{X}$. The Smart approach allows for  withstanding all known attacks even if the column scrambler matrix $ \mathbf{P}$ over the base field $\mathbb{F}_{q}$.

\end{abstract}

\section{Introduction}
McEliece \cite{McEliece} has introduced the first code-based public-key cryptosystem (PKC). The system is based on Goppa codes in the Hamming metric, which is connected to the hardness of the general decoding problem. It is a strong cryptosystem but the size of a public key is too large (500 000 bits) for practical implementations to be efficient.

Neiderreiter \cite{Niederreiter} has introduced a new PKC based on a family of Generalized Reed-Solomon codes; its public key size is less than the McEliece cryptosystem, but still large for practical application.

Also, Gabidulin Paramonov and Trietakov have proposed a new public key cryptosystem, which is now called the GPT cryptosystem, based on \emph{rank} error correcting codes in \cite{GPT1991, Gabidulin:1995/2}. The GPT cryptosystem has two advantages over McEliece's Cryptosystem. Firstly, it is more robust against decoding attacks than McEliece's Cryptosystem; secondly, the key size of the GPT is much smaller and  more useful in terms of practical applications  than McEliece's cryptosystem.

Rank codes are well structured. Subsequently in a series of works, Gibson \cite{Gibson:1995, Gibson:1996} developed attacks that break the GPT system for public keys of about $5$ Kbits. The Gibson's attacks   are efficient for practical values of parameters $n\leq 30$, where  $n$ is the length of rank code with the field $\mathbb{F}_{2^{N}}$ as an alphabet.

Several proposals  of the GPT PKC were introduced to withstand Gibson's attacks \cite{GabOuriv2000, ColumnScrambler2003}. One proposal is to use a rectangular row scramble matrix instead of a square matrix. The proposal allows  working with subcodes of the rank codes which have much more complicated structure. Another proposal exploits a modification of Maximum Rank Distance (MRD) codes where the concept of a \emph{column} scramble matrix was also introduced. A new variant, called reducible rank codes, is also implemented to modify the GPT cryptosystem \cite{ReducibleRankCodes2002, HighWeightErrors2005}. All these variants withstand Gibson's attack.\\

Recently, R. Overbeck \cite{Overbeck2005,Overbeck2006}, and \cite{Overbeck2008} has proposed  new attacks, which  are more effective than any of Gibson's attacks. His method is based on two factors : a) a column scrambler \emph{ P } that is defined over the \emph{base field} , and b) the unsuitable choice of a distortion matrix \emph{ X }.          However, Overbeck managed to break many instances of the GPT cryptosystem based on the general and developed ideas of Gibson.

Kshevetskiy in \cite{Kshevetskiy2006} suggested a secure approach towards the choice of parameters for avoiding Overbeck's attacks based on suitable choice of the distortion matrix X. Independently, Loidreau in \cite{Loidreau:2010} proposed similar method. Gabidulin  \cite{Gab2008} has offered a new approach called the Advanced approach, which makes the cryptographer define a proper column scrambler matrix over the extension field without violating the standard mode of the PKC. The Advanced approach allows the decryption of the authorised party, and prevents an unauthorized party from breaking the system by means of any known attacks.The two approaches withstand Overbeck and Gibson's attacks.

Recently, we have presented another variant of the GPT public key cryptosystem \cite{rashwan}, based on a proper choice of column scrambler matrix over the extension field. This variant, which we call the Instrumental approach, is secure against all known attacks.

In this paper, we introduce a new approach called the Smart approach, which is based on a proper choice of the distortion matrix \emph{ X }. The Smart approach allows for  withstanding all known attacks even if the column scrambler matrix $ \mathbf{P}$ over the base field $\mathbb{F}_{q}$.

The rest of this paper is structured as follows. Section 2 gives a short introduction to rank codes.  Section 3 describes the GPT cryptosystems. Section 4 discusses the Overbeck's attacks. Section 5 presents the Smart approach of GPT PKC cryptosystem with two examples. Finally, section 6 concludes the paper with some remarks.

\section{Rank codes}\label{sec:RankCodes}
\noindent Let us introduce the basic notion of rank codes
\cite{GPT1991}, \cite{Gab1985}. Let $\mathbb{F}_q$ be a
finite field of $q$ elements and let $\mathbb{F}_{q^N}$ be an
extension field of degree $N$. Let $\mathbf{x} = (x_1, x_2, \dots , x_n)$ be
a vector with coordinates in $\mathbb{F}_{q^N}$. \\
The Rank norm of $\textbf{x}$ is defined as the maximal number of
$\emph{x}_i$, which are linearly independent over the base field
$\mathbb{F}_q$ and is denoted $\mathrm{Rk}(\mathbf{x} \mid \mathbb{F}_q)$. \\
Similarly, for a matrix $\emph{M }$ with entries in
$\mathbb{F}_{q^N}$, the columns rank is defined as the maximal
number of columns, which are linearly independent over the base field $\mathbb{F}_q$,
and is denoted $\mathrm{Rk_{col}}(M |\mathbb{F}_q)$.\\
We distinguish two ranks of the matrix:
\begin{enumerate}
\item The usual rank of matrix $M$ over $\mathbb{F}_{q^N}$ -- $\mathrm{Rk}(M
\mid\mathbb{F}_{q^N})$.

\item  The column rank of a matrix $M$ over the base field
$\mathbb{F}_{q}$ -- $\mathrm{Rk_{col}}(M \mid\mathbb{F}_q)$.
\end{enumerate}
The column rank of the matrix $\emph{M }$ depends on the field. In particular,
$\mathrm{Rk_{col}}(M \mid\mathbb{F}_q)\geq \mathrm{Rk_{col}}(M |\mathbb{F}_{q^N})$\\
The Rank distance between $ \mathbf{x}$ and $ \mathbf{y}$ is defined
as the rank norm of the difference $ \mathbf{x-y}$: $d( \mathbf{x, y})
=\mathrm{Rk_{col}}( \mathbf{x-y} ~\mid ~ \mathbb{F}_q)$.\\

Any linear $(n,k,d)$ code  $ \mathcal{C}\subset\mathbb{F}^n_{q^N}$ fulfils the Singleton-style bound \cite{Gab1985} for the rank distance:
\begin{equation}\label{eq30}
    Nk\le Nn-(d-1)\max\{N,n\}.
\end{equation}

A code $\mathcal{C}$ reaching that bound is called a Maximal Rank Distance (MRD) code.

The theory of optimal MRD (Maximal Rank Distance) codes is given in
\cite{Gab1985}.

The notation $g[i] := g^{q^{i ~\mathrm{mod} ~n} }$ means the ${i}$-th
Frobenius power of $g$. It allows to consider both positive and negative Frobenius powers $i$.

For $n\le N$, a generator matrix $\mathbf{G}_k$ of a $(n,k,d)$ MRD code is
defined by a matrix of the following form:
\begin{equation}\label{eq1}
\mathbf{G}_{k}= 
\begin{bmatrix}
g_1 & g_2 & \dots & g_n\\
g_1^{[1]} & g_2^{[1]} & \dots & g_n^{[1]}\\
\vdots & \vdots & \ddots & \vdots\\
g_1^{[k-1]} & g_2^{[k-1]} & \dots & g_n^{[k-1]}
\end{bmatrix}
\end{equation}
where $g_1,g_2,\ldots,g_n$ are any set of elements of the extension
field $\mathbb{F}_{q^N}$ which are linearly independent over the
base field $\mathbb{F}_{q}$.\\
A code with the generator matrix \eqref{eq1}
is referred to as $(n, k, d)$ code, where $n$ is code length, $k$ is
the number of information symbols, $d$ is code distance. For MRD
codes, $d = n -k +1$. Let $ \mathbf{m}= (m_1,m_2, \dots,m_k )$ be an
information vector of dimension $k$. The corresponding code vector
is the $n$-vector
\begin{displaymath}
\mathbf{g}( \mathbf{m}) =  \mathbf{mG}_k .
\end{displaymath}
If $\mathbf{y} = \mathbf{g}(\mathbf{m}) + \mathbf{e}$ and $\mathrm{Rk}(\mathbf{e}) = s \leq t = \frac{d-1}{ 2}$ , then the
information vector $\mathbf{m}$ can be recovered uniquely from $\mathbf{y}$ by some
decoding algorithm. There exist fast decoding algorithms for MRD
codes \cite{Gab1985}, \cite{gab92}. A decoding
procedure requires elements of the $(n - k)\times n$ parity check
matrix $ \mathbf{H}$ such that $ \mathbf{G}_k \mathbf{H}^T = 0$. For
decoding, the matrix $ \mathbf{H}$ should be of the form
\begin{equation}\label{eq2}
\mathbf{H}=
\begin{bmatrix}
h_1 & h_2 & \dots & h_n\\
h_1^{[1]} & h_2^{[1]} & \dots & h_n^{[1]}\\
\vdots & \vdots & \ddots & \vdots\\
h_1^{[d-2]} & h_2^{[d-2]} & \dots & h_n^{[d-2]}
\end{bmatrix},
\end{equation}
where elements $h_1,h_2,\dots,h_n$ are in the extension field
$\mathbb{F}_{q^N}$ and are linearly independent over the base field
$\mathbb{F}_{q}$.

The optimal code has the following design parameters: code length $n \leq N$; dimension $k = n - d + 1 $, rank code distance $d = n - k +1 $.

\section{The GPT Cryptosystem}\label{sec:GPTcryptosystem}

\noindent\textbf{Description of the standard GPT cryptosystem.}\\
The GPT cryptosystem is described as follows:\\
\textbf{Plaintext}: A Plaintext is any $k$-vector
$\mathbf{m}=(m_1,m_2,\dots, m_k)$, $m_s\in \mathbb{F}_{q^N},\,~s=1,2,\ldots,k$.\\
In previous works, different representations of
the public key are given. All of them can be reduced to the
following form. \\
\textbf{The Public key} is a $k\times(n+t_1)$ generator matrix
\begin{equation}\label{eq3}
\mathbf{G}_{pub}= \mathbf{S}\begin{bmatrix}
                              \mathbf{X} & \mathbf{G}_{k} \\
                            \end{bmatrix} \mathbf{P}.
 \end{equation}
Let us explain roles of the factors.
\begin{itemize}
\item The main matrix $\mathbf{G}_k$ is given by \ref{eq1}.
It is used to correct rank errors. Errors of rank not greater than
$\frac{n-k}{2}$ can be corrected.
\item A matrix $\mathbf{S}$ is a row scrambler. This matrix is a non singular square matrix of order $k$ over $\mathbb{F}_{q^N}$.
\item A matrix $\mathbf{X}$ is a distortion $(k\times t_1)$ matrix over $\mathbb{F}_{q^N}$ with \textit{full} column rank $\mathrm{Rk_{col}}(X \mid\mathbb{F}_q)=t_1$ and rank
$\mathrm{Rk}(\mathbf{X} \mid\mathbb{F}_{q^N})=t_X, ~ t_X\leq t_1$. The matrix
$\begin{bmatrix}\mathbf{X} & \mathbf{G}_{k} \end{bmatrix}$
has full column rank $\mathrm{Rk_{col}}(\begin{bmatrix}
                              \mathbf{X} & \mathbf{G}_{k} \\
                            \end{bmatrix}\mid\mathbb{F}_q)=n+t_1$.
\item A matrix $\mathbf{P}$ is a square \emph{column scramble} matrix of order $(t_1+n)$ over $\mathbb{F}_{q}$.
\item $t_1+n$ may be greater than $N$, but $n\leq N$.
\end{itemize}
\textbf{The Private keys} are matrices
$ \mathbf{S},~ \mathbf{G}_k,~ \mathbf{X},~  \mathbf{P}$ separately and
(explicitly) a fast decoding algorithm of an MRD code. Note also,
that the matrix $ \mathbf{X}$ is not used to decrypt a ciphertext and
can be deleted after calculating the Public key.\\
\textbf{Encryption}: Let $\mathbf{m} = (m_1,m_2, \dots ,m_k )$ be a
plaintext. The corresponding ciphertext is given by
\begin{equation}\label{eq4}
\mathbf{c}= \mathbf{mG}_{\mathrm{pub}}+ \mathbf{e}= \mathbf{mS}\begin{bmatrix}
                              \mathbf{X} & \mathbf{G}_{k}                     \end{bmatrix} \mathbf{P}+ \mathbf{e},
 \end{equation}
where $ \mathbf{e}$ is an artificial vector of errors of rank $t_2$
or less. It is assumed that $t_1+t_2\leq t=\lfloor\frac{n-k}{2}\rfloor $\\

\textbf{Decryption}: The legitimate receiver upon
receiving $ \mathbf{c}$ calculates
\begin{displaymath}
 \mathbf{c}^{'}=(c_1^{'},c_2^{'},\ldots,c_{t_1+n}^{'})=
 \end{displaymath}
\begin{displaymath}
  \mathbf{c} \mathbf{P}^{-1} = \mathbf{mS}\begin{bmatrix}
                              \mathbf{X} & \mathbf{G}_{k} \\
                            \end{bmatrix}
+ \mathbf{e} \mathbf{P}^{-1}
 \end{displaymath}

Then from $ \mathbf{c}^{'}$ he extracts the subvector
\begin{equation}
 \mathbf{c}^{''}=(c_{t_1+1}^{'},c_{t_1+2}^{'},\ldots,c_{t_1+n}^{'})= \mathbf{mSG}_k
 + \mathbf{e}^{''},
\end{equation}
\noindent where $e^{''}$ is the subvector of $ \mathbf{eP}^{-1}$.
Then the legitimate receiver applies the fast decoding algorithm to
correct the error $ \mathbf{e}^{''}$, extracts $ \mathbf{mS}$ and
recovers $m$ as $ \mathbf{m}=( \mathbf{mS}) \mathbf{S}^{-1}$.\\
In this system, the size of the public key  is $V = k( t_1 + n )N$ bits,
and the information rate is $R = \frac{k}{t_1 + n}$.

\section{Overbeck's Attack}\label{sec:OverbecksAttack}

In \cite{Overbeck2005, Overbeck2006}, and \cite{Overbeck2008},  new attacks are proposed on the GPT PKC described in the form of \ref{eq3}. It is claimed, that similar
attacks can be proposed on all the variants of GPT PKC.

We recall briefly this attack.\\
We need some notations.\\
For $x\in \mathbb{F}_{q^N}$ let $\sigma (x)=x^q$ be the Frobenius
automorphism.\\
For the matrix $ \mathbf{T} = (t_{ij})$ over $\mathbb{F}_{q^N}$, let
$\sigma ( \mathbf{T})=(\sigma(t_{ij}))=(t_{ij}^q)$. \\
For any integer $s$, let $\sigma^s ( \mathbf{T})=\sigma(\sigma^{s-1}
( \mathbf{T}))$. \\
It is clear that $\sigma^N=\sigma$.  Thus the inverse exists
$\sigma^{-1}=\sigma^{N-1}$. \\
The following simple properties if $\sigma$ are useful:
\begin{itemize}
\item $\sigma(a + b) = \sigma(a)+\sigma(b)$.
\item  $\sigma(ab) = \sigma(a)\sigma(b)$.
\item In general, for matrices $\sigma( \mathbf{T})\neq \mathbf{T}$.
\item If  $\mathbf{P}$ is a matrix over the \emph{base} field $\mathbb{F}_q$, then
$\sigma( \mathbf{P}) =  \mathbf{P}$.
\end{itemize}
\textbf{Description of Overbeck's attack:} To break a system, a
cryptanalyst constructs from the public key
$ \mathbf{G}_{\mathrm{pub}}= \mathbf{S}\begin{bmatrix}
                              \mathbf{X} & \mathbf{G}_{k}                            \end{bmatrix}
 \mathbf{P}$ the \emph{extended} public key $\mathbf{G}_{\mathrm{ext,pub}}$ as follows:
\small
\begin{displaymath}
 \mathbf{G}_{\mathrm{ext,pub}}=
\left\|
\begin{matrix}
 \mathbf{G}_{\mathrm{pub}}\\
\sigma( \mathbf{G}_{\mathrm{pub}})\\
\dots\\
\sigma^u( \mathbf{G}_{\mathrm{pub}})\\
\end{matrix}
\right\|=
\end{displaymath}
\begin{equation}\label{eq.ext}
 \left\|
\begin{matrix}
 \mathbf{S} & \begin{bmatrix}
                              \mathbf{X}~~~~~ & ~~~~~\mathbf{G}_{k}                            \end{bmatrix}&   \mathbf{P}\\
\sigma( \mathbf{S})& \begin{bmatrix}
                              \sigma(\mathbf{X}) ~ & ~ \sigma(\mathbf{G}_{k})                            \end{bmatrix}&
 \mathbf{P}\\
\dots & \dots\dots~\dots~~~~~ &  \dots\\
\sigma^u( \mathbf{S})& \begin{bmatrix}
                              \sigma^u(\mathbf{X}) & \sigma^u(\mathbf{G}_{k})                            \end{bmatrix}&
 \mathbf{P}\\
\end{matrix}
\right\|.
\end{equation}
\normalsize
The property that $\sigma(\mathbf{P}) = \mathbf{P}$, if $\mathbf{P}$ is a matrix over the \emph{base} field $\mathbb{F}_q$, is used in \eqref{eq.ext}.

Rewrite this matrix as \small
\begin{equation}\label{eq5}
 \mathbf{G}_{\mathrm{ext,pub}} =  \mathbf{S}_{\mathrm{ext}}
\begin{bmatrix}
  \mathbf{X}_{\mathrm{ext}} & \mathbf{G}_{\mathrm{ext}}
\end{bmatrix}\mathbf{P},
\end{equation}\normalsize
where \small
\begin{equation}\label{eq:extended}
\begin{array}{c}
  \mathbf{S}_{\mathrm{ext}}=  \mathrm{Diag}\begin{bmatrix}
                                               \mathbf{S} & \sigma(\mathbf{S}) & \dots & \sigma^u(\mathbf{S})                                             \end{bmatrix}
   \\[3mm]
  \mathbf{X}_{\mathrm{ext}}=\begin{bmatrix}
                              \mathbf{X} \\
                              \sigma(\mathbf{X}) \\
                              \vdots \\
                              \sigma^u(\mathbf{X}) \\
                            \end{bmatrix},
  \quad \mathbf{G}_{\mathrm{ext}}=\begin{bmatrix}
                              \mathbf{G}_k \\
                              \sigma(\mathbf{G}_k) \\
                              \vdots \\
                              \sigma^u(\mathbf{G}_k) \\
                            \end{bmatrix}.

\end{array}
\end{equation}\normalsize
 Choose \small
\begin{equation}
u = n - k - 1 .
\end{equation}\normalsize

For a $k\times t_1$ matrix $\mathbf{X}$,
let $\mathbf{X}_1$ 
be the $(k-1)\times t_1$ matrix, obtained from $ \mathbf{X}$ by
deleting the \emph{last} row. Similarly, let $\mathbf{X}_2$
be the $(k-1)\times t_1$ matrix, obtained from $ \mathbf{X}$ by
deleting the \emph{first} row.

Define a linear mapping $T: \mathbb{F}_{q^N}^{k\times t_1}\rightarrow\mathbb{F}_{q^N}^{(k-1)\times t_1} $ by the rule:
if $\mathbf{X}\in \mathbb{F}_{q^N}^{k\times t_1}$, then \small
$
T(\mathbf{X})=\mathbf{Y}= \sigma(\mathbf{X}_1)-\mathbf{X}_2.
$ \normalsize
Let \footnotesize
\begin{equation}
 \mathbf{Y}_{\mathrm{ext}}=
\begin{bmatrix}
 \mathbf{Y}&
\sigma( \mathbf{Y})&
\sigma^2( \mathbf{Y})&
\dots&
\sigma^{u-1}( \mathbf{Y})
\end{bmatrix}^{\top}
\end{equation}
\normalsize
Using this and other suitable transformations of rows, one can rewrite for analysis
\eqref{eq5} and \eqref{eq:extended} in the form \footnotesize
\begin{equation}\label{eq11}
\tilde{ \mathbf{G}}_{\mathrm{pub,ext}}=\tilde{ \mathbf{S}}_{\mathrm{ext}}
\begin{bmatrix}
 \mathbf{Z}&|& \mathbf{G}_{n-1}\\
 \mathbf{Y}_{\mathrm{ext}}&|&0\\
\end{bmatrix}
\mathbf{P}
\end{equation}\normalsize
where $ \mathbf{G}_{n-1}$ is the generator matrix of the $(n,n-1,2)$
MRD code.

Let us try to find a solution $\mathbf{u}$ of the system \footnotesize
\begin{equation}\label{eq:Main}
	\tilde{ \mathbf{S}}_{ext}
\begin{bmatrix}
 \mathbf{Z}&|& \mathbf{G}_{n-1}\\
 \mathbf{Y}_{ext}&|&0\\
\end{bmatrix}
 \mathbf{P} \mathbf{u}^T= \mathbf{0},
\end{equation}\normalsize
where $\mathbf{u}$ is a vector-row over the extension field $\mathbb{F}_{q^N}$ of length $t_1+n$. Represent the vector $\mathbf{P} \mathbf{u}^T$ as
\[
\mathbf{P} \mathbf{u}^T=\begin{bmatrix} \mathbf{y} & \mathbf{h}\end{bmatrix}^T,
\]
where the subvector $\mathbf{y}$ has length $t_1$ and $\mathbf{h}$ has length $n$. Then the system \eqref{eq:Main} is equivalent to the following system:\footnotesize
\begin{align}
\mathbf{Z}\mathbf{y}^T+\mathbf{G}_{n-1}\mathbf{h}^T=\mathbf{0},\label{eq:Main1}\\
	\mathbf{Y}_{ext}\mathbf{y}^T=\mathbf{0}.\label{eq:Main2}
\end{align}\normalsize
Assume that the next condition is valid: \footnotesize
\begin{equation}\label{eq12}
 \mathrm{Rk}(\mathbf{Y}_{ext}
|\mathbb{F}_{q^N})=t_1.
\end{equation} \normalsize
Then the equation  \eqref{eq:Main2} has only the trivial solution $\mathbf{y}^T=\mathbf{0}$. The  equation \eqref{eq:Main1} becomes \footnotesize
\begin{equation}\label{eq:Main3}
\mathbf{G}_{n-1}\mathbf{h}^T=\mathbf{0}.	
\end{equation}\normalsize
It allows to find the first row of the parity check matrix for the code with the generator
matrix \eqref{eq11} (see,\cite{Overbeck2005,Overbeck2006}, and \cite{Overbeck2008}, for details). Hence this solution breaks a GPT cryptosystem in polynomial time.  The Overbeck's attack requires $O((n+t_1)^3)$ operation over
$\mathbb{F}_{q^N}$ since all the steps of the attack have at most
cubic complexity on $n+t_1$.

\section{Smart approach}\label{sec:Smart Approach}
To withstand  Overbeck's attack, the cryptographer should choose the matrix $\mathbf{X}$ in such a manner that\footnotesize
\begin{equation}\label{eq13}
\mathrm{Rk}( \mathbf{Y}_{ext}\mid
\mathbb{F}_{q^N})=t_1-a,
\end{equation}\normalsize
where $a \geq 2$. In this case, the system \eqref{eq:Main2} has $q^{aN}$  solutions $\mathbf{y}^T$. Hence the exhaustive search over $\mathbf{y}^T$ is needed. The work function has order
$O(q^{aN}(n+t_1)^3)$ and Overback's attack fails.

One method to provide the condition \eqref{eq13} is proposed in \cite{Kshevetskiy2006, Loidreau:2010}.  Choose the matrix $\mathbf{X}$ over the extension field $\mathbb{F}_{q^N}$ in such a manner that the following conditions are satisfied:\footnotesize
 \begin{equation}\label{eq:RankX}
 \begin{array}{lclcl}
 t_1&=&\mathrm{Rk_{col}}(\mathbf{X}\mid \mathbb{F}_{q})&>& n-k.\\
	r_X&=&\mathrm{Rk}(\mathbf{X}\mid \mathbb{F}_{q^N})&=&\left\lfloor  \frac{t_1-a}{n-k}\right\rfloor\le k.
	\end{array}
\end{equation}\normalsize
Overbeck's attack is exponential on $a$ and has the minimum  complexity at least $O\left(q^{aN}(n+t_1)^3\right)$.

We propose an alternative Smart approach. The point is to choose the matrix $\mathbf{X}$ in such a manner that the corresponding matrix $\mathbf{Y}=T(\mathbf{X})$  has column rank $\mathrm{Rk}(\mathbf{Y}\mid \mathbb{F}_q)$ not greater than $t_1-a,\,a\ge 2$.

The following result is evident.
\begin{lemma}\label{lem:RkY}
 If $\mathrm{Rk}(\mathbf{Y}\mid \mathbb{F}_q)=s$, then $\mathrm{Rk}(\mathbf{Y}_{\mathrm{ext}}\mid \mathbb{F}_q)=s$.
\end{lemma}
\begin{corollary}\label{cor:1}\footnotesize
$\mathrm{Rk}(\mathbf{Y}_{\mathrm{ext}}\mid \mathbb{F}_{q^N})\le \mathrm{Rk}(\mathbf{Y}_{\mathrm{ext}}\mid \mathbb{F}_q)=s=\mathrm{Rk}(\mathbf{Y}\mid \mathbb{F}_q)$.\normalsize
\end{corollary}

\paragraph{\textbf{The simple case}} Let a matrix $\mathbf{X}$ be of the following form:\footnotesize
\begin{equation}\label{eq:MatrixXspecial}
	\mathbf{X}=\begin{bmatrix}\mathbf{m}\\
	\mathbf{m}^{[1]}\\
	\vdots\\
	 \mathbf{m}^{[k-1]}\end{bmatrix}+\begin{bmatrix}\mathbf{0}\\
	\mathbf{s}_1\\
	\vdots\\
	 \mathbf{s}_{k-1}\end{bmatrix}.
\end{equation}\normalsize
Here $\mathbf{m}$ is a random vector over the extension field $\mathbb{F}_{q^N}$ with full column rank $t_1$ and vectors $\mathbf{s}_i,\;i=1,\dots,k-1,$ are random vectors over the \emph{base} field  $\mathbb{F}_{q}$ such that the  matrix\footnotesize
\[
\begin{bmatrix}\mathbf{0}&
	\mathbf{s}_1&
	\dots&
	 \mathbf{s}_{k-1}\end{bmatrix}^{\top}
\]\normalsize
has rank $t_1-a$. Then the matrix $\mathbf{Y}=T(\mathbf{X})$ has the form\footnotesize
\begin{equation}\label{eq:MatrixYspecial}
    \mathbf{Y}=\begin{bmatrix}-\mathbf{s}_1&
	\mathbf{s}_1-\mathbf{s}_2&
	\dots&
	 \mathbf{s}_{k-1}-\mathbf{s}_k\end{bmatrix}^{\top}.
\end{equation}\normalsize
This matrix is a matrix over the \emph{base} field $\mathbb{F}_{q}$ and has rank $t_1-a$ too. It follows that\footnotesize
\begin{equation}\label{eq:sigmaY}
  \sigma(\mathbf{Y})=\begin{bmatrix}\sigma(-\mathbf{s}_1)\\
	\sigma(\mathbf{s}_1-\mathbf{s}_2)\\
	\vdots\\
	 \sigma(\mathbf{s}_{k-1}-\mathbf{s}_k)\end{bmatrix}=\begin{bmatrix}-\mathbf{s}_1\\
	\mathbf{s}_1-\mathbf{s}_2\\
	\vdots\\
	 \mathbf{s}_{k-1}-\mathbf{s}_k\end{bmatrix}=\mathbf{Y}.
\end{equation}\normalsize
Hence\footnotesize
 \begin{equation}
 \mathbf{Y}_{ext}=
\begin{bmatrix}
 \mathbf{Y}\\
\sigma( \mathbf{Y})\\
\dots\\
\sigma^{u-1}( \mathbf{Y})\\
\end{bmatrix}
=\begin{bmatrix}
 \mathbf{Y}\\
 \mathbf{Y}\\
\dots\\
 \mathbf{Y}\\
\end{bmatrix}.
\end{equation}\normalsize
Therefore 
$
\mathrm{Rk}(\mathbf{Y}_{ext}\mid \mathbb{F}_{q^N})=\mathrm{Rk}(\mathbf{Y}\mid \mathbb{F}_{q^N})=t_1-a,
$
and the condition \eqref{eq13} is  satisfied.

As in the previous case, the proposed Smart approach shows that Overbeck's attack is exponential on $a$ and has the bit complexity at least $O\left(q^{aN}(n+t_1)^3\right)$.

It has been shown  that the Smart  approach presented above is secure against all known attacks including the recent attack presented by Overbeck in \cite {Overbeck2008}.
\small
\begin{example}
Let $n=8,~ k=4,~N=8,~ t=5,~t_1=4, ~q=2,~ a=2$\\
Let the extension field $\mathbb{F}_{2^8}$ be defined by the primitive polynomial
$
r(x)=1+x^2+x^3+x^4+x^8,
$
and let $\alpha$ be a primitive element of the field. Choose the matrix $ \mathbf{X}$ as in \eqref{eq:MatrixXspecial}. A vector $\mathbf{m}$ of full column rank $t_1=4$ is defined as
$
\mathbf{m}=\begin{bmatrix}
             \alpha^3 & \alpha^5 & \alpha^6 & \alpha^2 
           \end{bmatrix}.
$
Choose vectors $\mathbf{s}_1,\mathbf{s}_2,\mathbf{s}_3$ as $\mathbf{s}_1=\begin{bmatrix}
                                                                           1 & 1 & 0 & 0 
                                                                         \end{bmatrix}
$, $\mathbf{s}_2=\begin{bmatrix}
                                                                           1 & 1 & 1 & 1 
                                                                         \end{bmatrix}
$, $\mathbf{s}_3=\begin{bmatrix}
                                                                           0 & 0 & 1 & 1 
                                                                         \end{bmatrix}.
$
Then we obtain \footnotesize
\begin{equation}\label{eq:Xexample}
\begin{array}{rl}
    \mathbf{X}=&\begin{bmatrix}
             \alpha^3 & \alpha^5 & \alpha^6 & \alpha^2 \\
             \alpha^6 & \alpha^{10} & \alpha^{12} & \alpha^4 \\
             \alpha^{12} & \alpha^{20} & \alpha^{24} & \alpha^8 \\
             \alpha^{24} & \alpha^{40} & \alpha^{48} & \alpha^{16} \\
           \end{bmatrix}+\begin{bmatrix}
             0 & 0 & 0 & 0 \\
             1 & 1 & 0 & 0 \\
             1 & 1 & 1 & 1\\
             0 & 0 & 1 & 1 \\
           \end{bmatrix}= \\  [3mm]

           &\begin{bmatrix}
             \alpha^3 & \alpha^5 & \alpha^6 & \alpha^2 \\
             \alpha^6+1 & \alpha^{10}+1 & \alpha^{12} & \alpha^4 \\
         \alpha^{12}+1 & \alpha^{20}+1 & \alpha^{24}+1 & \alpha^8+1 \\
             \alpha^{24} & \alpha^{40} & \alpha^{48}+1 & \alpha^{16}+1 \\
           \end{bmatrix}
           \end{array}.
\end{equation}
\small
The corresponding matrix $\mathbf{Y}$ is as follows:\footnotesize
\begin{equation}\label{eq:Yexample}
    \mathbf{Y}=\begin{bmatrix}
             1 & 1 & 0 & 0 \\
             0 & 0 & 1 & 1 \\
             1 & 1 & 0 & 0 \\
                \end{bmatrix}.
\end{equation}
\small
It has rank $t_1-a=2$.
The attack is exponential on $a$ and has the bit complexity at least
$O(q^{aN}(n+t_1)^3)=O(2^{37}$ bite operations.
\end{example}\normalsize
\paragraph{\textbf{The general case}} Let $\mathbf{X}$ be a matrix consisting of $a$ Frobenius-type columns and $t_1-a$ non-Frobenius columns. A column $\mathbf{w}$ is called Frobenius-type if it has the form $\mathbf{w}=\begin{pmatrix}                                               w & w^{[1]} & \dots & w^{[k-1]}\end{pmatrix}^{\top}$. It is clear that $T(\mathbf{w})=\mathbf{0}$. Hence the matrix $\mathbf{Y}=T(\mathbf{X})$ will have $a$ all zero columns and column rank $t_1-a$ and by Corollary \ref{cor:1} the matrix $\mathbf{Y}_{\mathrm{ext}}$ has rank not greater than $t_1-a$. The result is valid also if suitable linear combinations of non-Frobenius columns are added to Frobenius-type columns.
\small \begin{example}
In conditions of the previous example, let matrix $\mathbf{X}$ be as follows:\small
\[
\mathbf{X}=\begin{bmatrix}
             \alpha^3+\alpha^6 & \alpha^5+ \alpha^2 & \alpha^6 & \alpha^2 \\
             \alpha^6+\alpha^{12} & \alpha^{10}+\alpha^5 & \alpha^{12} & \alpha^5 \\
             \alpha^{12}+\alpha^{12} & \alpha^{20}+\alpha^5 & \alpha^{12} & \alpha^5 \\
             \alpha^{24}+\alpha^{12} & \alpha^{40}+ \alpha^2 & \alpha^{12} & \alpha^{2} \\
           \end{bmatrix}.
\]
The third column is added to the first Frobenius-type, and the fourth is added to the second Frobenius-type, so $a=2$. Column rank of $\mathbf{X}$ is $t_1=4$. The corresponding matrix $\mathbf{Y}=T(\mathbf{X})$ is of the form:\small
\[
\mathbf{Y}=\begin{bmatrix}
             0 & \alpha^{4}+\alpha^5  & 0 & \alpha^{4}+\alpha^5 \\
             \alpha^{24}+\alpha^{12} & \alpha^{4}+\alpha^5 & \alpha^{24}+\alpha^{12} & \alpha^{4}+\alpha^5 \\
             \alpha^{24}+\alpha^{12} & \alpha^{10}+\alpha^5 & \alpha^{24}+\alpha^{12} & \alpha^{10}+\alpha^5 \\
           \end{bmatrix}.
\]\normalsize
It has rank $t_1-a=2$.
\end{example}

In general, Overbeck's attack fails when $aN\ge 60$.
\section{Conclusion}

We have introduced the Smart approach as a technique of withstanding Overbeck's attack on the GPT Public key cryptosystem,  which is  based on rank codes.

It is shown that proper choice of the distortion matrix $\mathbf{X}$ over the extension field
$\mathbb{F}_{q^{N}}$ allows the decryption by  the authorized party and prevents the unauthorized party from
breaking the system by means of any known attacks.

\end{document}